\documentclass[]{spie}  

\usepackage{amsmath,amssymb}
\usepackage[pdftex]{graphicx,color}
\usepackage{url}
\usepackage{subfigure}
\title{The Good, the Bad, and the Ugly: three different approaches to break their watermarking system}

\author{G. Le Guelvouit\supit{a}, T. Furon\supit{b}, and F. Cayre\supit{c} \skiplinehalf
            \supit{a}France Telecom R\&D, Cesson S\'evign\'e, France\\
            \supit{b}INRIA, Rennes, France\\
            \supit{c}LIS, Grenoble, France
           }

\authorinfo{Further information: (Send correspondence to T.F.): teddy.furon@irisa.fr}

\begin{document}
\maketitle

\begin{abstract}
The Good is Blondie, a wandering gunman with a strong personal sense of honor. The Bad is Angel Eyes, a sadistic hitman who always hits his mark. The Ugly is Tuco, a Mexican bandit who's always only looking out for himself. Against the backdrop of the BOWS contest, they search for a watermark in gold buried in three images. Each knows only a portion of the gold's exact location, so for the moment they're dependent on each other. However, none are particularly inclined to share\ldots
\end{abstract}


\section{Once upon a time in the watermarking community}\label{sect:intro}

This paper presents the work of three researchers participating in the BOWS contest. The tone of the paper is largely humorous and ironic. We hope nobody will take it badly or personally. There are also many quotations from spaghetti western movies. However, section~\ref{sec:Good} contains some more serious stuff dealing with optimal attack for spread spectrum based watermarking schemes.

	\subsection{{\em ``For a fistful of dollars''} and a digital camera}

The BOWS contest (Break Our Watermarking System) was organized by the watermarking virtual lab (WAVILA) of the European Network of Excellence (NoE) ECRYPT. It ran for three months between December and March 2006. A secret zero-bit watermarking technique was used to watermark three unknown $512 \times 512$ images.
Their watermarked versions were posted on the website. The watermark detector was online. The contestants had to modify these images and test if the detector could still find the presence of the invisible watermark. The winner of the contest was the one who could remove the watermark while providing the best quality for the three images. The PSNR between the watermarked and the attacked version was the measure of quality.
The prize was \$300 and a digital camera. Indeed, it is quite a pity that the prize of a contest organised by an European NoE was not given in Euro: 300~Euros would have been more money\ldots

	\subsection{\em ``My name is nobody''}

During SPIE Electronic Imaging, January 2005, I asked one of the three members of the BOWS steering committee to give a status of BOWS. At the time, the committee was quite unhappy because they had only 1000 image submissions after a month, and, especially, there were not enough good hackers. Nine persons had hacked the first image, but nobody succeeded on the three images.

The man was torn with sadness, repentance, and disappointment. The challenge was too hard. The lack of winner could mean that his challenge was fake and as serious as a bad media operation to promote watermarking.

As I committed some papers about watermarking and security, he finally asked me why I was not in the mood of participating. My position about the challenge was clear. I heard once one of the best cryptographers in the world saying that, times to times, people asked him to break a cipher text. His answer is always the same: {\em ``I am not a bounty hunter, I am a researcher.''} His point is that decrypting a ciphertext, when nothing is known, looks like more a DIY activity than research work. He prefers challenges where the algorithm of crypto-system is public and where breaking the system has a scientific value.       
Following the spirit of this `godfather', I found that the BOWS challenge was not correctly set up to motivate research activity.
It would have been better to disclose the watermarking technique. The steering committee member disagreed and said that the challenge is open to anybody, not only watermarking researchers.
Now that the challenge is over, I have changed my mind and apology. Good lessons are learned from this contest (see Sec.~\ref{sec:lessons}). 

I ended the discussion saying that I would not participate to the challenge, however I would engage two gunmen, Blondie and Tuco, to seek this fortune. I did not tell him how Angel Eyes would motivate these two desperados. 

A very important information is that, at that time, the number of submissions was limited.

\section{The role of Angel Eyes Sentenza: social engineering}

There were at least two flaws in the organisation of the contest. 

	\subsection{\em ``In this world, there's two kinds of people''}

{\em ``You see, in this world there's two kinds of people, my friend: those with loaded guns and those who dig. You dig.''}
Exactly, and in this contest, there's two kinds of hackers: those who knew the would be secret watermarking technique and the others.

A first mistake is that the members of the steering committee first announced their will to organizing a watermarking challenge. I guess they were testing people reactions, wondering whether they would have been interested in participating, or trying to get support from the ECRYPT network. There has been a lot of discussions on how the challenge must be set up. Once the steering committee had clearer ideas on this topic, they had to search for a watermarking technique.    

From Angel Eyes's point of view, it means that the hacker was aware that there had been a quest for this watermarking technique, a long time before the challenge actually started. Actually, the hacker started to work before the challenge was opened, wandering which watermarking technique he would choose, if he were a steering committee member.

\begin{description}
	\item[No industrial product.] The steering committee had planned to disclose the watermarking technique in a second step of the contest. No company would agree on letting its technology being used in the contest under this condition. Moreover, what a bad advertising operation if the challenge is easily broken. Does it ring a bell? Yes exactly, SDMI! Thus the technique has been invented by someone who doesn't care: a pure academic researcher!
	\item[Something made in Europe.] As the contest is an ECRYPT event, i.e. organized and supported by Europeans, then, this more or less implies that the watermarking technique is European.
	\item[No synchronization needed.] As the PSNR is the measure of quality, the hackers cannot {\it a priori} use a geometrical attack (no rotation, no scaling, no sheering, etc.), unless the committee makes the mistakes to use cover images invariant to one of these attacks. But, this was not the case. This argument actually largely broadens our scope of research, while discarding techniques robust to geometric attacks. If the technique was robust to geometric attack, the committee would have been glad to strengthen the challenge in the sense that geometric attack would have been allowed.
	\item[Something working.] The steering committee will be successful if the challenge becomes famous and well considered by the community. The committee miserably fails if the technique can be too easily hacked. They need something known for being robust at least against common image processing giving a PSNR higher than 30~dB.
	\item[Something ready.] They look for a real watermarking technique not a watermarking scheme working on Gaussian vectors. The steering committee took about three months to  select a technique and develop the server infrastructure. There is a little chance that they build something from scratch, even if these three men are known to be dangerous and efficient gunfighters.
\end{description}         

We will not dare listing the potential teams having such a watermarking technique available, as we would forget people, putting us in an awful diplomatic situation. However, if the reader knows the watermarking community, he will agree that this list is not that long.

To tell the truth, the list reduced to one item one day of September 2005. I was attending IWDW 2005, and I noticed that Prof. Barni and G. Do\"err had a lot of discussion during the coffee breaks. They were softly speaking slightly away from the attendees. They did not see I was approaching them. I heard Prof. Barni {\em ``When do you think it will be ready?''} before these two gentlemen realized I was there, they smiled, visibly annoyed that I surprised them. {\em ``Oh, how do you find Siena? A marvelous place, isn't it?''}. That's it. I was almost sure of the origin of the watermarking technique.

	\subsection{\em ``When you have to shoot, shoot, don't talk''}

This is where social engineering enters in the picture. Social engineering is the art of obtaining confidential information simply talking to people. The famous motto is \textit{users are the weak link} in security. There are numerous examples of these tricks, and also rules and policies to avoid such trivial hacks~\cite{SocialWikiPedia,SocialMitnick}.

At that time, we needed a confirmation: Gwena\"el Do\"err is in charge of developing the watermarking technique, and he is certainly implementing his famous algorithm co-invented with M.~Miller and I.~Cox~\cite{Miller2004:Applying}. Now, what do we know about this character? He is French, he loves wines, and he is always late. So, I took care of sitting just beside him at the social event of IWDW 2005, and started asking questions.

\begin{itemize}
	\item {\em ``What are you working on now?''} I knew this question was too straight. Gwena\"el avoided answering, saying he had difficulty coaching lazy students in the lab. Gwena\"el is a good guy, he dislikes saying lies, and I noticed that as he was not comfortable. This was the right moment to cleverly manipulate his natural human tendency to trust.
	\item {\em ``Mauro told me about BOWS,''} I said. A typical sentence having many possible meanings in case my intuition was wrong. Yet, if I was right to suspect Gwena\"el, it tells him that I am not a stranger in the organization, and I am coming from the boss authority so that he can trust me. Gwena\"el answered he was feeling bad about BOWS because he was pretty late.
	\item {\em ``Well, I told Mauro that I can help you, if needed.''} This is the perfect position in social hacking to pull information out of the victim: pretend you are here to help him.
	\item {\em ``So, where do you stand in the source code?''} I added, while being careful that his glass of wine was never empty. From his answers, I learned that my intuition was correct with respect with the watermarking technique~\cite{Miller2004:Applying}, that Gwena\"el used the same DCT coefficients to embed the watermark as in his article, and that the length of the sub-carriers was not yet decided. He was actually doing some experimental investigations on this topic, analyzing the trade-off between robustness and the probability of false alarm. Anyway, I was nice to propose help, but he will manage this on his own.
\end{itemize}

It was just that easy. However, we think that without social engineering, it was possible to guess the watermarking technique. The decomposition in $8 \times 8$ blocks is easily seen in the watermarked images, whereas they are in raw format. The watermark is robust to an valuemetric change (i.e. a change in contrast), such that a QIM based technique is out of suspicion. All these considerations really narrow the set of potential techniques. We were certainly not the only ones knowing the secret. When the watermarking technique had been officially revealed for the second round of the contest, very few researchers actually improved their scores. {\em ``In this world, there's two kinds of people\ldots''}

	\subsection{The church, the wood path and the strawberry} 

How many strawberries (image 1) and wood paths (image 2) are there in the world? Quite a lot. How many churches (image 3)? Certainly less. Each one being very specific and located in very different places. However, a church is not a random building. There exist construction types, styles that give clues on the period of time and the location of the building. Notice also the mountain in the background\ldots Buy your tickets and grab your camera, we finally located it!

Well, this is not true. We did not succeed. But, there is something much more easier to be done. If I were a steering committee member, which pictures should I use for the challenge? Obviously, I should be the only one possessing the original version of these pictures. So, it is certainly some personal pictures of one of the committee members, or pictures he used for testing his own watermarking techniques in his lab. In this case, he might have already published it in his article for illustration purpose. This would be an enormous flaw! But it is true! The reader should look at figure 3 page 64 in the proceedings of the $6^{\scriptsize \textrm{th}}$ Information Hiding workshop \cite{Abrardo}.

It is the same church, but it is not the same picture. First, this picture is watermarked by the watermarking technique proposed by A.~Abrardo and M.~Barni at an embedding distortion of 37.5~dB in PSNR~\cite{Abrardo}. Second, the image is scaled and the contrast is not the same. We manually synchronized the two images. Then, an exhaustive search  giving the highest PSNR is proceeded over the following parameters: horizontal offset, vertical offset, contrast, and scale factor. The result is a hacked picture at 37.35~dB.

\begin{figure}[htbp]
	\begin{center}
		\subfigure[Watermarked picture from the BOWS website]{\includegraphics[width=6cm]{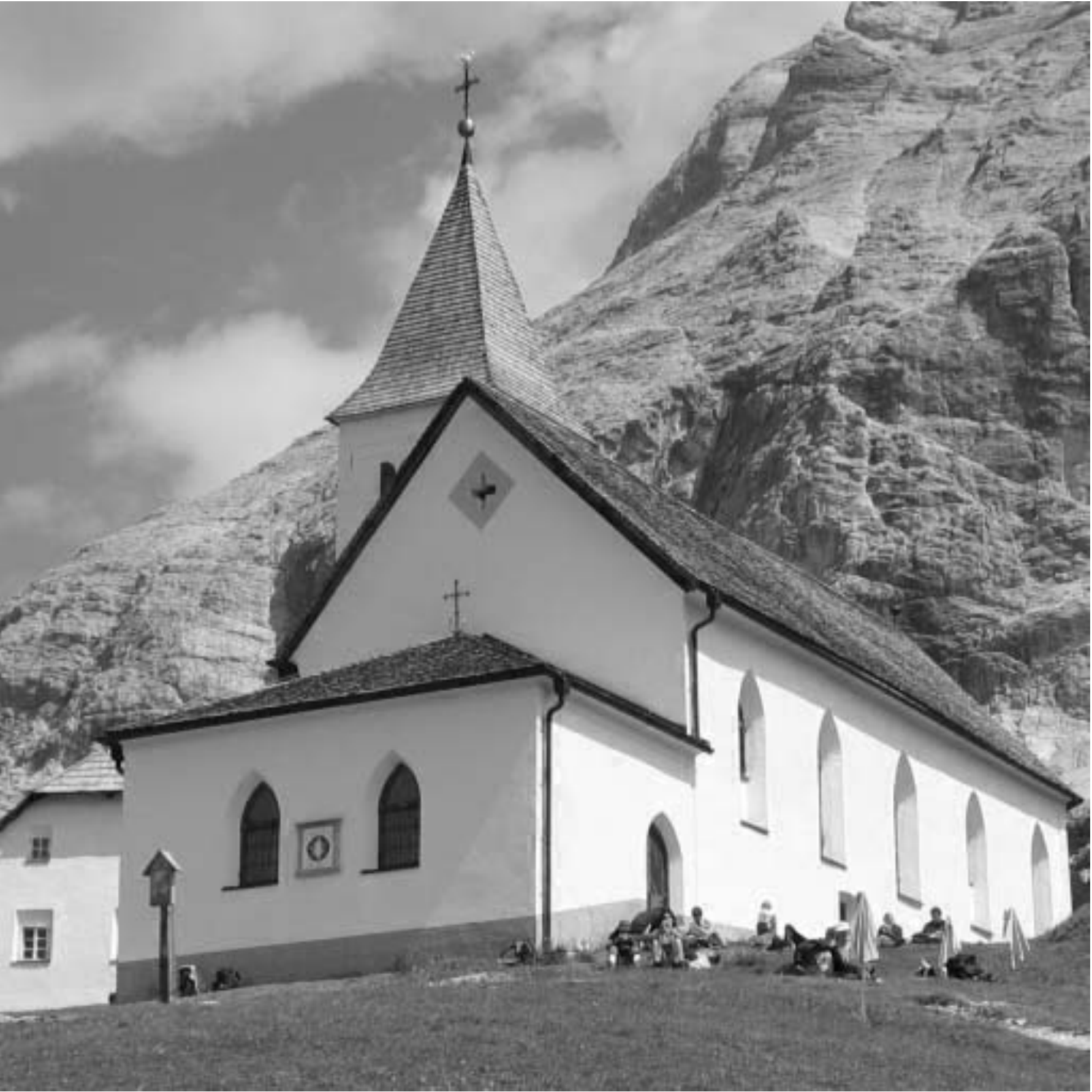}}
		\hspace{0.5cm}
		\subfigure[Picture captured from A.~Abrardo and M.~Barni paper~\cite{Abrardo}]{\framebox{\includegraphics[width=6cm]{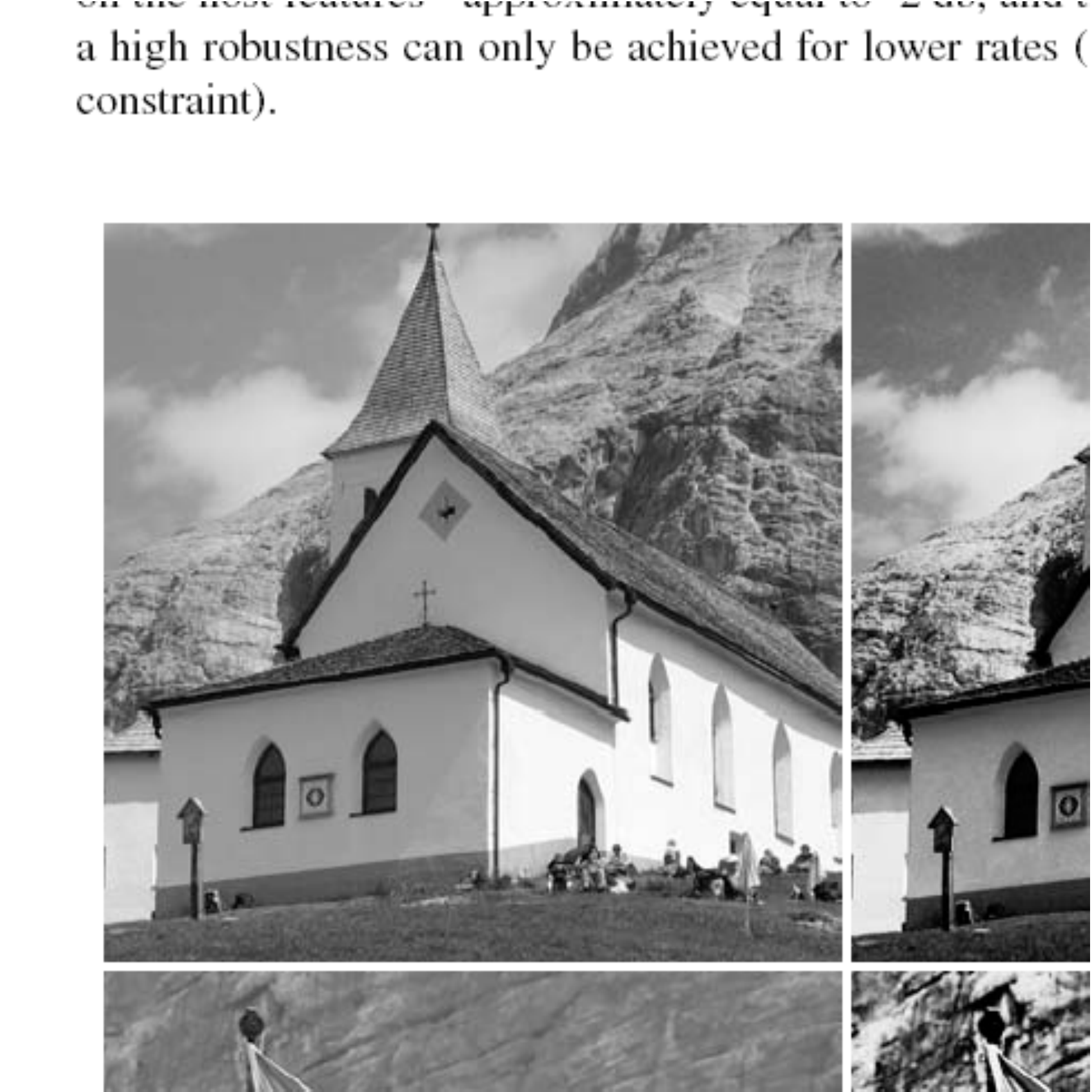}}}
		\caption{Two pictures of the same church.}
		\label{fig:church}
	\end{center}
\end{figure}

\section{Tuco, the Ugly}\label{sec:Ugly}  

Before explaining how ugly I was in attacking BOWS algorithm, I have to mention that Sentenza only hired me for \$100 and I later learned the true amount of the prize. At that time, I did not know of Blondie neither. I suspect Sentenza wanted to split the cash into three (more or less) equal parts, keeping the camera for himself\footnote{I was told he recently ran amok shooting churches in the French and Italian countryside while keeping buying indecent quantities of strawberries, even in December, which he reportedly exclusively eats in wood paths.}. Anyway, I decided to spend two days attacking the algorithm, just to prove him my good will and to distract me from my lazy students. 

	\subsection{Setting up}

Contrary to Sentenza, I made no questionable inference on the algorithm but relied on my personal belief that the algorithm was the one with this funny trellis\cite{Miller2004:Applying}. Since, from SDMI, we know it is weird to have such a challenge fashion, I thought the algorithm itself should be weird too. I'd better have used sooner the well-known beer-based variant hack\footnote{Which I successfully and repeatedly applied during the last IH workshop in Alexandria, VA, although for other purposes.} to make G. Do\"err speak:
I did not think the sub-carriers length would have changed from the paper (12 DCT coefficients). Shame on me.

As a respected authority in ugliness, I personnally would regard it as being very ugly to use interpreter-based tools. So I spent a few minutes writing the basic C routines for reading the pictures from scratch. And I am such an ugly and dyslexic person that I incidentally managed to swap my line/column variables. I started by attacking the strawberry image. I was so confident that I did not even write the PGM output routines for inspecting the visual quality of my attack. I obtained a watermark-free strawberry image with a PSNR higher than 34~dB. After 2 hours of adding ugly noise to my hand-made DCT coefficients, I was happy. Actually, I was so happy that I eventually wrote my PGM output routines to admire the result of my brilliant attack. It was just perfect, except the displeasing effect of producing a diagonal-flipped image. I therefore decided with magnanimity to no longer submit diagonally-flipped images to the detector. This strawberry image is so diagonally oriented that such a rapid success could certainly not be so easily achievable with the two other images. So I concluded that geometrical resynchronization may sometimes be helpful. But it obviously does not accommodate the chosen algorithm very well. To respect the weird spirit of the West, I decided to turn my attack into something weirder than noise addition. 

	\subsection{WUA: Weird and Ugly Attack} 

This attack could in no way be applied as is to any image. I manually tuned several parameters to accommodate each attack until the detector said it could not find the watermark anymore. Basically, I used two complementary strategies: 
\begin{itemize}
	\item setting all DCT coefficients that were at user-defined distance from the AC coefficient to zero, 
	\item random shuffling of $(0,1)$, $(1,0)$, $(1,1)$, $(2,0)$, $(2,1)$, $(1,2)$, $(2,2)$ and $(0,2)$ DCT coefficients. 
\end{itemize}

\begin{figure}[htbp]
	\begin{center}
		\subfigure[Strawberry ($30.39$~dB)]{\includegraphics[width=5cm]{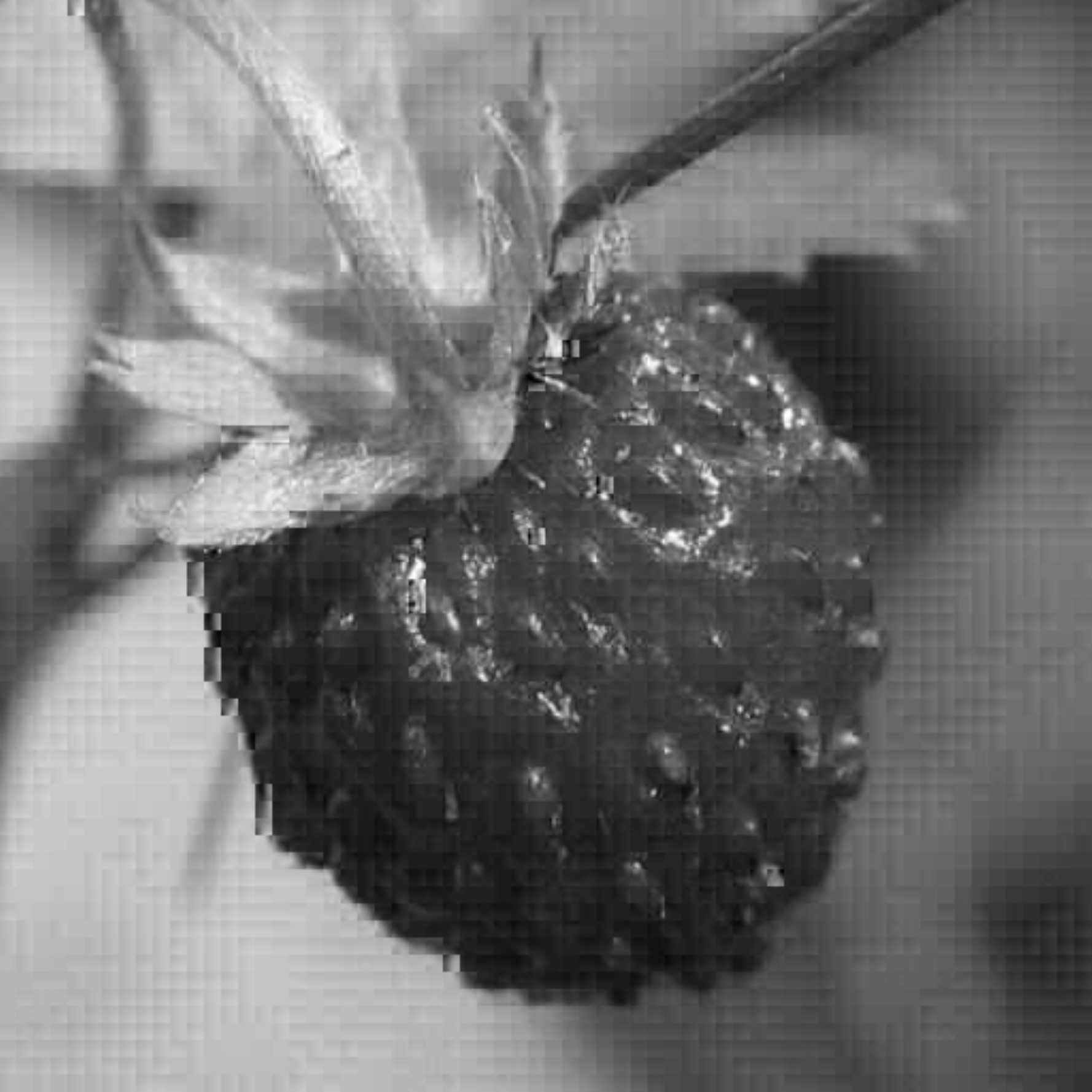}}
		\hspace{0.5cm}
		\subfigure[Wood path ($31.21$~dB)]{\includegraphics[width=5cm]{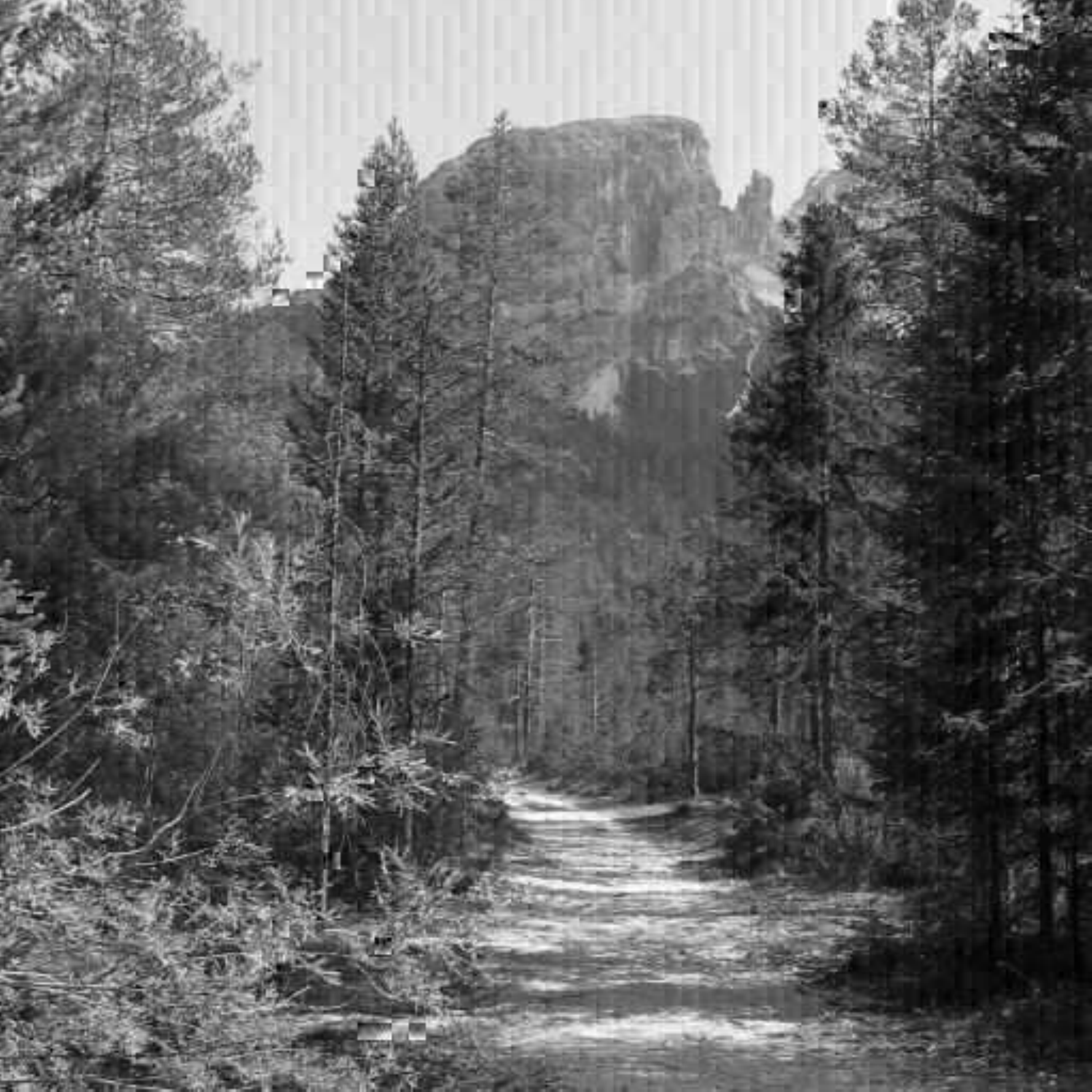}}
		\hspace{0.5cm}
		\subfigure[Church ($30.22$~dB)]{\includegraphics[width=5cm]{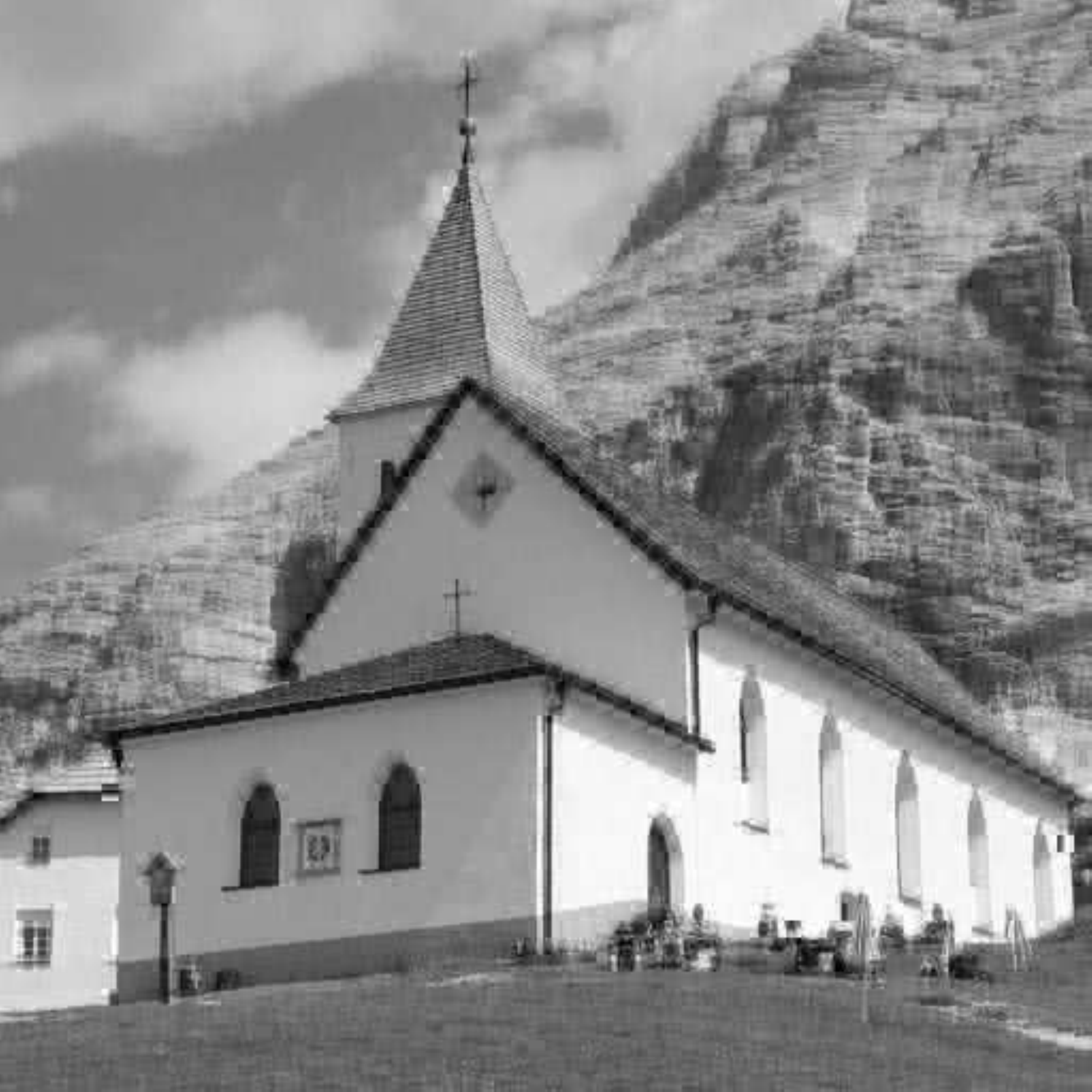}}
		\caption{Successfully attacked images with the WUA attack depicted by Tuco (mean PSNR of $30.59$~dB).}
	\label{fig:weirdandugly}
	\end{center}
\end{figure}

What I call random shuffling is somewhat tricky: I first performed an index-based sorting of every $(x,y)$ coefficients (i.e. I sorted every $(1,2)$ coefficients, every $(0,1)$ coefficients, etc. and considered these families of coefficients independently). I tuned the power of my attack with an integer $A$. I then randomly swapped the under-attack $(x,y)$ DCT coefficient with another one that was in the same index-ordered array of $(x,y)$ coefficients, but at maximum index position difference being $A$. As we all know, some seeds are better than others and I tried a few dozen for each image. Recall that I decided to spend no more than two days with this challenge. I may have also added some noise on some images. My goal was not to have beautiful attacked images (see results on Fig.~\ref{fig:weirdandugly}): I just wanted to attack the detector PSNR-wise. I eventually obtained attacked images with strong blocking effect that are very annoying for the eye. Quoting Coco Chanel: {\em ``Art is ugly things that become beautiful and fashion is beautiful things that become ugly''}, I was unable to do art in this challenge, I just did challenge-fashion victims with three images. But beware: when it comes to fashion, Blondie is awesome. 

\section{Blondie, the Good}\label{sec:Good}

This is the Good part of the paper, in the sense that I (Blondie) did not need to spy on respectable researchers and to make people drunk before shooting. As seen in first section, one can deduce some hints concerning the watermarking technique used thanks to simple tests. A close look at the three watermarked images shows blocking artifacts, like JPEG compression. So, embedding is certainly made in the DCT domain, using $8 \times 8$ blocks. A couple of attempts also shows that the watermark does not suffer from valuemetric scaling. It is a typical property of spread spectrum based techniques. I then started working with those two hypotheses.

	\subsection{A theoretical optimal attack}
	
The use of game theory to study possible interactions between embedders and attackers is a major advance in watermarking theory. This leads to the design of new embedding techniques, but also to new attack strategies. We will minimize detector's gains to maximize the efficiency of our attack\footnote{This part is only an adaptation of previous work~\cite{gleugleu} for the contest.}.\\

First, we must quantify the gain. As seen in literature, watermarking can be seen as communication over a channel. We use the channel capacity to quantify the gain. Blondie's work is to minimize this gain. Let us denote $\mathbf x$ the host signal. In the context of the challenge and considering our hypotheses, it is a vector of $m = 512 \times 512$ samples issued of a DCT. Each sample is modeled by $\mathbf X[i] \sim \mathcal N(0, \textrm{var}(\mathbf X[i]))$. Spread spectrum watermarking consists in modulate a set of pseudo-random carriers by the bits to be embedded. Let $\mathbf m \in \mathbb R^n$ be those bits and $\mathbf G \in \{ -1 ; +1 \}^{m \times n}$ be the carriers. The watermarked vector is thus defined as 
\begin{equation}
	\mathbf y[i] = \mathbf x[i] + \mathbf w[i] = \mathbf x[i] + \mathbf a[i] \sum_{j=1}^n \mathbf m[j] \times \mathbf G[i][j] \textrm{,}
\end{equation}
where $\mathbf a[i] \geq 0$ is a weighting factor. In this study, we consider SAWGN attacks, i.e. a scaling factor and the addition of Gaussian noise. After attack, the received vector is denoted as 
\begin{equation}
	\mathbf y'[i] = \mathbf s[i] \times \left( 
		\mathbf y[i] + \mathbf z[i] 
	\right) \textrm{,}
\end{equation}
where $\mathbf s[i] \geq 0$	 is the scaling factor and $\mathbf z[i]$ is modeled by $\mathbf Z[i] \sim \mathcal N(0, \textrm{var}(\mathbf Z[i]))$. It has been demonstrated in numerous papers that the use of pseudo-random carriers defines a Gaussian channel with side information available at the encoder. Considering SAWGN attacks, the capacity of such a channel is 
\begin{eqnarray}
	C = \frac 1 2 \log_2 \left[
		1 + \sum_{i=1}^m \textrm{snr}(i)
	\right] \textrm{~where~} \textrm{snr}(i) &=& \frac {\mathbf a[i]^2}{\textrm{var}(\mathbf Z[i])} 
	\textrm{~if~} \mathbf s[i]>0 \\ 
	&=& 0 \textrm{~otherwise.}
\end{eqnarray}
Blondie's quest is to minimize $C$, that is to minimize each $\textrm{snr}(i)$. \\
		
Since the contest allows a minimum PSNR of 30~dB between watermarked and attacked images, we introduce a distortion measure based of mean square error. Let $D$ be this measure:
\begin{eqnarray}
	D &=& \mathbb E \left[
		\sum_{i=1}^m \left( 
			\mathbf y[i] - \mathbf y'[i]
		\right)^2 
	\right] \nonumber \\
	&=& \sum_{i=1}^m \mathbf y[i]^2 (1 - \mathbf s[i])^2 + \mathbf s[i]^2 \times \textrm{var}(\mathbf Z[i]) \textrm .
\end{eqnarray}
We then use a Langrangian formulation to take into account both the gain and the distortion:
\begin{equation}
	J_\lambda = \sum_{i=1}^m \textrm{snr}(i) + \lambda D 
	= \sum_{i=1}^m \left[
		\frac {\mathbf a[i]^2}{\textrm{var}(\mathbf Z[i])} + \lambda \left[
			\mathbf y[i]^2 (1 - \mathbf s[i])^2 + \mathbf s[i]^2 \times \textrm{var}(\mathbf Z[i]) 
		\right] 
	\right] \textrm . 
\end{equation}
Solving this minimization shows that the optimal form of $\mathbf s[i]$ looks like a Wiener filter, and that the optimal variance of added noise $\mathbf Z[i]$ is function of $\mathbf a[i]$ (watermark energy) and $\mathbf y[i]$ (sample to be attacked): 
\begin{eqnarray}
	\mathbf s[i] &=& \lambda \frac{\mathbf y[i]^2}{\mathbf y[i]^2 + \mathbf z[i]^2}  \textrm ,\\
	\textrm{var}(\mathbf Z[i]) &=& \mathbf a[i] \left[ 
		\frac{\lambda^{\frac 3 2} \mathbf y[i]^2 - \mathbf a[i]}{\lambda^{\frac 3 2} \mathbf y[i]^2}
	\right] \textrm . \label{eq:var}
\end{eqnarray}
Nevertheless, since $\textrm{var}(\mathbf Z[i]) \geq 0$, Eqn.~(\ref{eq:var}) is valid only if $\mathbf a[i] \leq \lambda^{\frac 3 2} \mathbf y[i]^2$. Otherwise, no noise must be added and $\mathbf s[i]$ must be set to zero, i.e. the attack consists in erasing the considered sample. So Blondie has two choices: erasing or filtering after the addition of Gaussian noise. But the optimal parameters we found depend on the embedded watermark energy. Since Blondie does not precisely know this value, he must estimate it prior to the attack.
	
	\subsection{A much less theoretical estimation of embedding strategy}
	
\begin{figure}[htbp]
	\begin{center}
		\subfigure[Strawberry, using optimal attack ($34.95$~dB)]{\includegraphics[width=5cm]{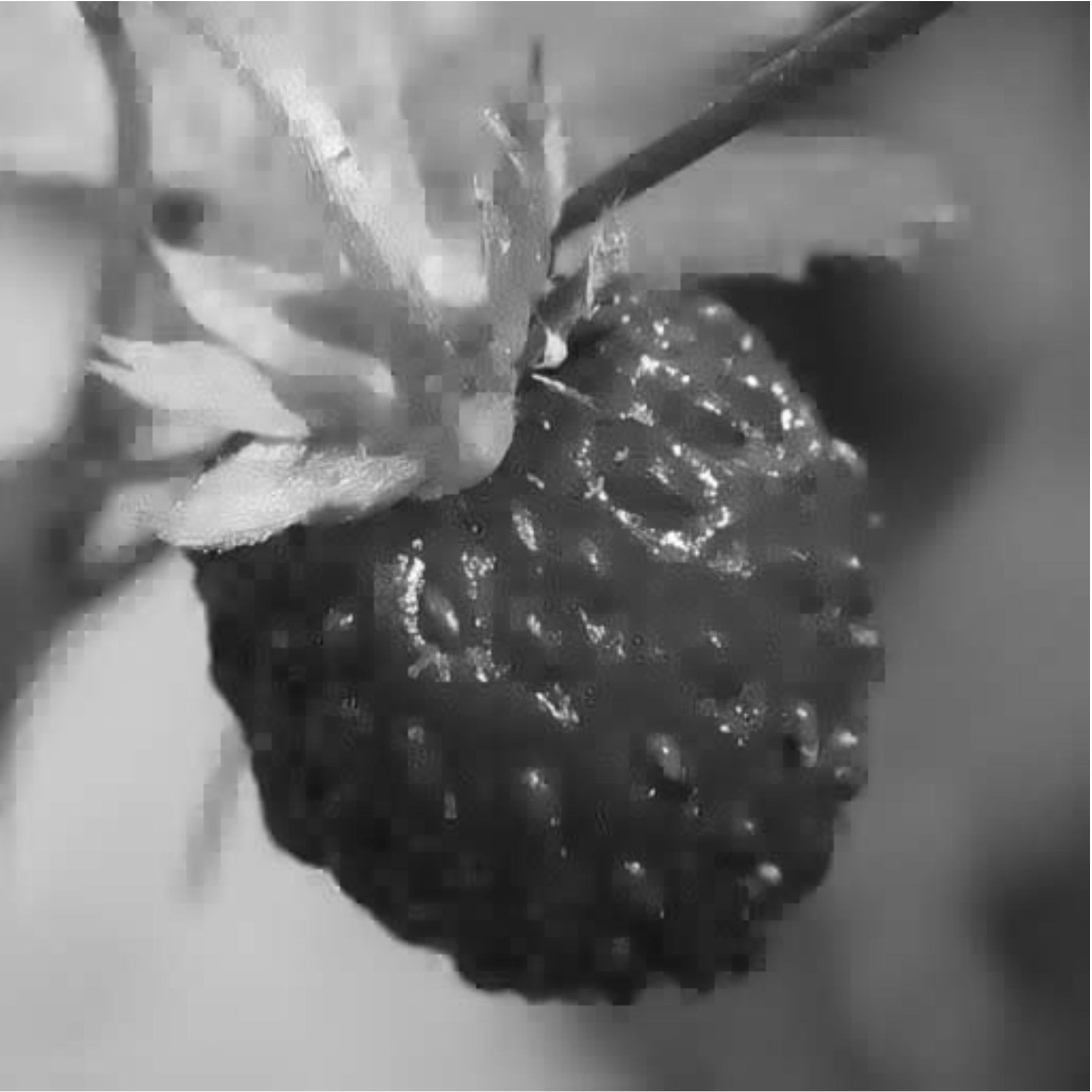}}
		\hspace{0.5cm}
		\subfigure[Wood-path, using optimal attack ($34.04$~dB)]{\includegraphics[width=5cm]{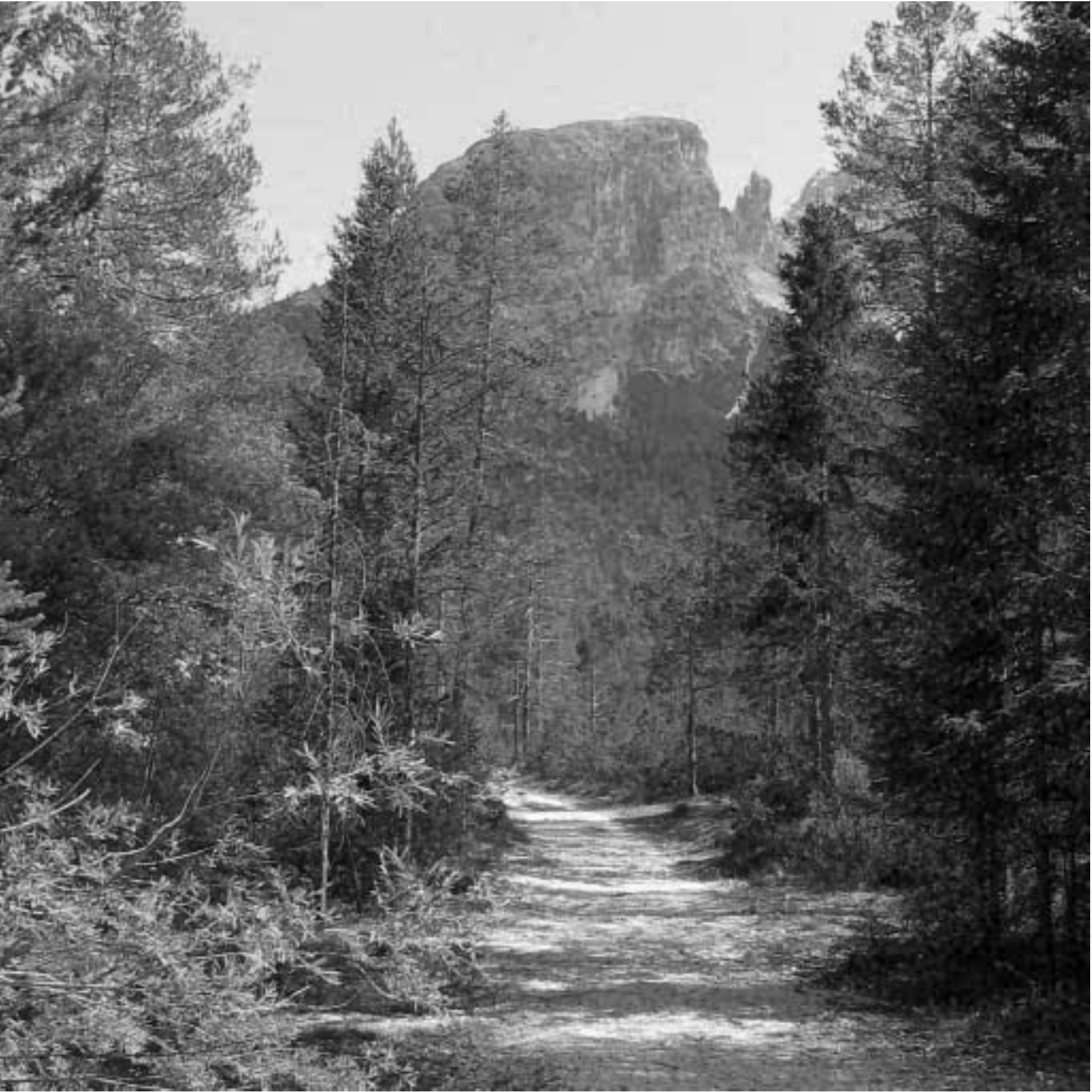}}
		\hspace{0.5cm}
		\subfigure[Church, using Abrardo~{\it et al.} paper ($37.35$~dB)]{\includegraphics[width=5cm]{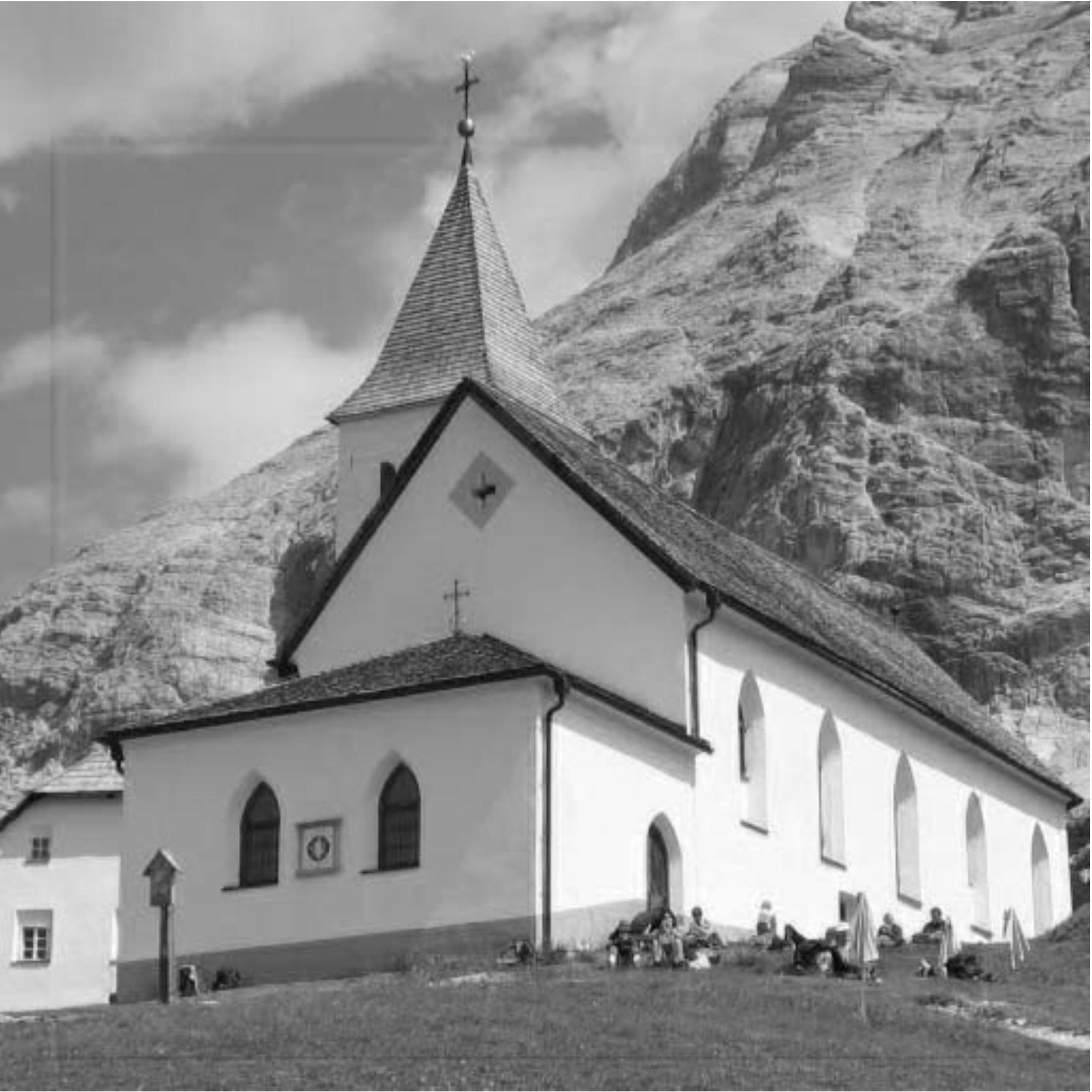}}
		\caption{Attacked images (mean PSNR of $35.24$~dB).}
		\label{fig:attacked}
	\end{center}
\end{figure}

Angel Eyes told Blondie his conviction about the use of Miller~{\it et al.} watermarking technique~\cite{Miller2004:Applying}. After reading once more the corresponding paper, Blondie focuses only on the twelve DCT samples used in the paper for embedding\footnote{Furthermore, some attempts showed that strong attacks of the others DCT samples are not successful.}, thus $\mathbf a[i]$ is estimated to be $0$ for 52 others samples. For the remaining samples, Blondie assumed the use of a perceptual factor. Watermark energy was supposed to be similar to Watson's perceptual factor (also used in Miller's experiments), i.e $\mathbf a[i] = k \times \textrm{watson}(i)$. Better results were obtained tuning parameters $k$ and $\lambda$. This finally leaded to the first two attacked images from Fig.~\ref{fig:attacked}.
	
\section{Lessons from the BOWS contest}\label{sec:lessons}

We did not win the challenge. We were third ($35.24$~dB). However, the second contestant, John Earl ($37.27$~dB), registered his scores the $16^{\scriptsize \textrm{th}}$ of March whereas the official deadline of the first part of the challenge was $15^{\scriptsize \textrm{th}}$ of March. The winner, S.~Craver and his team ($39.22$~dB), used an oracle attack, which was at the beginning of the challenge forbidden by the time constraint on the online detector. We elaborated our strategy before the steering committee changed this rule. This implies a first lesson: attacks based on strong theoretical foundation work great in practice, i.e. the worst case attack (or optimal attack) and the oracle attack (when possible).

Reading the report of A.~Westfeld on the second phase of the contest, it appears that focusing all the attack distortion on only three DCT coefficients removes the watermark at a PSNR of 40 dB. We think that this is due to a misconception of the watermarking technique. The challenge was about zero-bit watermarking (i.e. watermark detection) whereas the selected technique embeds several bits. The detector was indeed a watermark decoder which outputs `Yes' when the decoded binary message equals the reference message shared by the embedder and the decoder. The trade-off between robustness and payload more or less implies that a watermark decoder has a Message Error Rate greater than the probability of misdetection of a watermark detector. This is especially true as the hidden bits are embedded via a time division multiple access (TDMA). Hence, focusing the attack on the region where a single bit is hidden easily causes a decoding error. Of course, the DCT coefficients have certainly been permutated by an interleaver before embedding, such that this focused attack is not practical\ldots except when an oracle attack is allowed.

\bibliographystyle{spiebib}   
\bibliography{EI-2007}   

\begin{thebibliography}{1}

\bibitem{SocialWikiPedia}
{Wikipedia contributors}, ``Social engineering (security),'' {\em Wikipedia,
  the free encyclopedia,
  \url{http://en.wikipedia.org/wiki/Social_engineering_%28computer_security%29%
}} , July~2006.

\bibitem{SocialMitnick}
K.~D. Mitnick, W.~L. Simon, and S.~Wozniak, {\em The art of deception:
  controlling the human element of security}, John Wiley \& Sons, 2002.

\bibitem{Miller2004:Applying}
M.~L. Miller, G.~Do{\"e}rr, and I.~J. Cox, ``Applying informed coding and
  informed embedding to design a robust, high capacity watermark,'' {\em IEEE
  Trans. Image Processing}~{\bf 13}, pp.~792--807, June~2004.

\bibitem{Abrardo}
A.~Abrardo and M.~Barni, ``Fixed-distortion orthogonal dirty paper coding for
  perceptual still image watermarking,'' {\em Proc. of the $6^{\scriptsize
  \textrm{th}}$ Information Hiding Workshop}~{\bf 3200}, pp.~52--66, May~2004.

\bibitem{gleugleu}
S.~Pateux and G.~{Le Guelvouit}, ``Practical watermarking scheme based on wide
  spread spectrum and game theory,'' {\em Signal Processing: Image
  Communication}~{\bf 18}, pp.~283--296, Apr.~2003.

\end{thebibliography}

\end{document}